%% file: main.tex
\documentclass[preprint]{imsart}

\usepackage{amsmath,amsthm,amssymb,amsfonts,fullpage,verbatim,bm,graphicx,enumerate,epstopdf,lscape,gensymb,appendix}
\usepackage[dvipsnames,usenames]{color}
\usepackage[bookmarks=false]{hyperref}
\usepackage{enumitem,indentfirst,pdflscape,geometry}
\setenumerate{listparindent=24pt}
\usepackage{moreverb}
\usepackage{natbib}
\usepackage{subcaption,caption}
\usepackage{booktabs}

\def\G{\mathcal G}

\def\bhat{\hat{\beta}}
\def\shat{\hat{s}}

\def\yv{\mathbf y}
\def\ev{\mathbf e}

\def\xv{\mathbf x}

\def\b{\beta}
\def\s{s}
\def\bhat{\hat{\b}}
\def\shat{\hat{\s}}
\def\df{\nu}

\def\voom{\emph{voom}}
\def\ash{\emph{ash}}

\def\qvalue{\emph{qvalue}}

\def\va{\emph{VL+ash}}
\def\vla{\emph{VL+eBayes+ash}}
\def\vlaa{\emph{VL+eBayes+ash.alpha=1}}
\def\vlpa{\emph{VL+pval2se+ash}}

\def\vl{\emph{VL+eBayes}}
\def\vlq{\emph{VL+eBayes+qvalue}}

\newtheorem{lemma}{Lemma}

\begin{document}
\begin{frontmatter}
\title{Empirical Bayes estimation of normal means, accounting for uncertainty in estimated standard errors\thanksref{T1}}
\runtitle{}
\thankstext{T1}{This work was supported by NIH grant HG002585 and by a grant from the Gordon and Betty Moore Foundation}

\begin{aug}
\author{\fnms{Mengyin} \snm{Lu} %\thanksref{t1}
\ead[label=e1]{mengyin@uchicago.edu}}
\and
\author{\fnms{Matthew} \snm{Stephens} %\thanksref{t1}
\ead[label=e2]{mstephens@uchicago.edu}}

\runauthor{Lu and Stephens}

\affiliation{University of Chicago\thanksmark{m1} }

\address{
%Address of the First and Second authors\\
%Usually a few lines long\\
\printead{e1}\\
\printead*{e2}\\
% \address{Address of the Third author\\
% Usually a few lines long\\
% Usually a few lines long\\
}
\end{aug}

\begin{abstract}
We consider Empirical Bayes (EB) estimation in the normal means problem, when the standard
deviations of the observations are not known precisely, but estimated with error -- which 
is almost always the case in practical applications. 
In classical statistics accounting for estimated standard errors usually involves
replacing a normal distribution with a $t$ distribution.
%More specifically, we consider EB estimation
%of means, $b_j$ ($j=1,\dots,p$), given observations $\bhat_j$ and estimated standard errors $\hat{s}_j$,
%that satisfy $(\bhat_j - \b_j)/\shat_j \sim t_{\df_j}$ for known degrees of freedom $\df_j$.
This suggests approaching this problem by replacing the normal assumption with
a $t$ assumption, leading to an ``EB $t$-means problem".
Here we show that an approach along these lines can indeed work, but only with some care.
Indeed, a naive application of this idea is flawed, and can perform poorly.
We suggest how this flaw can be remedied
by a two-stage procedure, which first performs EB shrinkage estimation of the standard errors 
and then solves an EB $t$-means problem. We give numerical results illustrating the
effectiveness of this remedy.
\end{abstract}

% \begin{keyword}[class=MSC]
% \kwd[Primary ]{60K35}
% \kwd{60K35}
% \kwd[; secondary ]{60K35}
% \end{keyword}


\end{frontmatter}

%\title{Empirical Bayes estimation of normal means, accounting for uncertainty in estimated standard errors}
%\date{}
%\maketitle

\section{Introduction}

We consider Empirical Bayes (EB) estimation in applications where we have observed estimates $\bhat_j$ ($j=1,\dots,p$) of a series of underlying ``effects" $\b_j$, with estimated standard
errors $\shat_j$. Our goal is to perform EB estimation for the effects
$\b_1,\dots,\b_p$, from the observations $\bhat_1,\dots,\bhat_p$, under standard normal theory assumptions, but {\it taking account of uncertainty in the standard errors}. 

If the standard errors of $\bhat_1,\dots,\bhat_p$ were known, rather than estimated, our problem would simply involve EB inference for the well-studied ``Normal means'' problem \citep[e.g.][]{johnstone2004}:
\begin{align} \label{eqn:bhatj}
\bhat_j | \b_j, \s_j & \sim N(\b_j, \s_j^2), \qquad j=1,\dots,p;\\ \label{eqn:bj}
\b_j | \s_j & \sim g_\b \in \G, \qquad j=1,\dots,p;
\end{align}
where $\s_j$ denotes the ``true'' standard error of $\bhat_j$, and $\G$ is some specified family of distributions. (The conditioning on $\s_j$ in \eqref{eqn:bj} makes explicit an assumption that the $\beta_j$ are independent and identically distributed from $g_\beta$, independent of $s_j$, an assumption we relax later.) Fitting this EB model involves first obtaining an estimate $\hat{g}_\b$ for $g_\b$ (e.g. by marginal maximum likelihood), and then computing the posterior distributions $p(\b_j | \bhat_j,\s_j, \hat{g}_\b)$. These posterior distributions
can be used to obtain both point and interval estimates of $\b_j$. And, if the family $\G$ involves sparse distributions (with a point mass on 0), then the posterior distributions can also be used to compute (local) false discovery rates \citep{Efron2004}, effectively providing an EB solution to the ``multiple testing" problem.  This EB normal means problem is well studied, and there exist flexible software implementations for solving it for a range of choices of $\G$ \citep[e.g.][]{Stephens2016}.
For example, methods
in \cite{Stephens2016} effectively solve this problem for $\G$ the set of all unimodal distributions (by exploiting the fact that any such
distribution can be approximated, to arbitrary accuracy, by a mixture of sufficiently many uniform distributions).

In classical statistics, the fact that standard errors are estimated
is usually dealt with by replacing normal distributions with $t$ distributions. Indeed, in the settings we consider here, we have 
\begin{equation}
(\bhat_j - \b_j)/s_j \sim N(0,1),
\end{equation}
and
\begin{equation} \label{eqn:t}
(\bhat_j - \b_j)/\shat_j \sim t_{\df_j},
\end{equation}
where $t_\df$ denotes the $t$ distribution on $\df$ degrees of freedom.
Expression \eqref{eqn:t} is routinely used in classical statistics to obtain confidence intervals for $\b_j$ and $p$ values testing $\b_j=0$.

From this it is tempting to replace the EB normal means problem \eqref{eqn:bhatj}-\eqref{eqn:bj} with what we call the ``EB $t$-means problem" (EBTM):
\begin{align} \label{eqn:bhatj-t}
\bhat_j | \b_j, \shat_j & \sim t_{\df_j}(\b_j, \shat_j) \\ \label{eqn:bj-t}
\b_j | \shat_j & \sim g_\b \in \mathcal{G},
\end{align}
where $t_\df(\mu, \sigma)$ denotes the generalized $t$ distribution on $\df$ degrees of freedom, with mean $\mu$ and scale parameter $\sigma$ (i.e. the distribution of $\mu + \sigma T$ when $T \sim t_\df$). While this EBTM problem is much less studied than the EB normal means problem, \cite{Stephens2016} also provides flexible software implementations solving the EBTM problem -- estimating $g$ and computing posterior distributions $p(\b_j | \bhat_j,\shat_j, \hat{g})$ -- for a range of choices of $\G$.

Unfortunately, there is a problem with this tempting naive approach: while
the EBTM problem \eqref{eqn:bhatj-t}-\eqref{eqn:bj-t} is well-defined and solvable, the standard theory that leads to \eqref{eqn:t} does not imply \eqref{eqn:bhatj-t}. The
reason is that in \eqref{eqn:t} $\shat_j$ is random, and not conditioned on,
and the unconditional expression does not imply a corresponding conditional one:
\begin{equation} \label{eqn:not-imply}
(\bhat_j - \b_j)/\shat_j \sim t_\df \nRightarrow (\bhat_j - \b_j)/\shat_j | \shat_j \sim t_\df.
\end{equation}
To give a simple explicit example of this: if $\bhat_j \sim N(0,1)$ and $\shat_j^2 \sim \chi_1^2$ then $\bhat_j/\shat_j \sim t_1$ but $\bhat_j /\shat_j | \shat_j \sim N(0,1/\shat_j^2)$.
Consequently \eqref{eqn:bhatj-t} does not hold in general,
and -- as we show later -- ignoring this can produce very unreliable inferences in practice.

In this paper we describe a simple solution to this problem. Our solution
involves EB analysis of the standard errors $\shat_j$ \citep{smyth2004limma}, which is already widely used in genomics applications -- indeed, currently much more
widely used than EB analysis of the effect estimates $\bhat_j$.
Our approach effectively combines the methods from \cite{smyth2004limma} with the methods for the EBTM problem from \cite{Stephens2016}.
We demonstrate empirically that, in contrast with 
the naive approach, this combined
approach can provide reliable inference.

\section{Methods}

Assume that, independently for $j=1,\dots,p$, we have
observed estimates $\bhat_j$ and corresponding (estimated) standard errors $\shat_j$, satisfying
\begin{equation} \label{eqn:a1}
p(\bhat_j,\shat_j | \b_j,\s_j) = p(\shat_j | \s_j) p(\bhat_j | \b_j, \s_j)  
\end{equation}
where
\begin{align} \label{eqn:a2}
\bhat_j | \b_j, \s_j & \sim N(\b_j, \s_j^2) \\  \label{eqn:a3}
\shat^2_j | \s_j & \sim \s^2_j \chi^2_\df/\df.
\end{align}
For example \eqref{eqn:a1}-\eqref{eqn:a3} hold if $\bhat_j,\shat_j$ are
the usual estimate of $\b_j$ and its standard error in a simple linear regression, $\yv_j = \xv \b_j + \ev_j$, where $\yv_j$ and $\xv$ are observed $n$-vectors and the residual errors $\ev_j \sim N(0,\sigma_j^2 I_n)$, with $s^2_j:=(\xv^T\xv)^{-1} \sigma_j^2$.

Our goal is to perform EB estimation for $\b_1,\dots,\b_p$ under the assumption \eqref{eqn:bj-t} that $\b_j | \shat_j \sim g_\b \in \G$. As noted in the Introduction, if \eqref{eqn:bhatj-t} held then this would be solved by methods for the EBTM problem in \cite{Stephens2016}. However, unfortunately \eqref{eqn:a1}-\eqref{eqn:a3} do not imply \eqref{eqn:bhatj-t} and so \eqref{eqn:bhatj-t} does not hold in general.

We now describe a simple solution to this problem, based on combining the EBTM methods in \cite{Stephens2016} with EB estimation for $\shat_j$ using the methods in \cite{smyth2004limma}. Specifically, \cite{smyth2004limma} combines the sampling distribution for $\shat_j$ \eqref{eqn:a3} with
an assumption that the true variances $\s^2_j$ come from an inverse-gamma distribution, which can be written:
\begin{equation} \label{eqn:seprior}
\s^{-2}_j \sim \s_0^{-2} \chi^2_{\df_0} / \df_0 \qquad (j=1,\dots,p)
\end{equation}
where $\s_0,\df_0$ are parameters to be estimated.  The EB approach in \cite{smyth2004limma} estimates $\s_0,\df_0$ from the observations $\shat_1,\dots,\shat_p$ (using a method of moments), and then bases inferences for $\s^2_j$ on its posterior distribution given these estimates, which is also an inverse-gamma distribution. Indeed, given 
$s_0$ and $\df_0$, the posterior can be written
\begin{equation} \label{eqn:spost}
\s^{-2}_j | \shat_j \sim \tilde{s}_j^{-2} \chi^2_{\tilde\df_j}/\tilde\df_j,
\end{equation}
where
\begin{align} \label{eqn:dftilde}
\tilde\df_j &:= \df_0 + \df_j \\ \label{eqn:stilde}
\tilde{s}_j^{2} &:=(\df_0 \s_0^2 + \df_j \shat_j^2)/(\df_0+\df_j).
\end{align}
In particular \cite{smyth2004limma} uses $\tilde{\s}_j^2$ -- which lies between $s_0^2$ and $\shat_j^2$ -- 
as a ``moderated" estimate of $\s^2_j$.

The key to our approach is the following simple Lemma.
\begin{lemma} \label{lemma}
Assuming \eqref{eqn:a1},\eqref{eqn:a2} and \eqref{eqn:spost} it follows that %\eqref{eqn:seprior}, and given $s_0,\df_0$, then
\begin{equation} \label{eqn:bhatj-t2}
\bhat_j | \b_j, \shat_j \sim t_{\tilde{\df}_j}(\b_j, \tilde{\s}_j).
\end{equation}
\end{lemma}
Thus, although \eqref{eqn:bhatj-t} does not hold in general, under
the assumptions \eqref{eqn:a1}-\eqref{eqn:seprior} (which imply \eqref{eqn:spost}) an analogous 
expression \eqref{eqn:bhatj-t2} {\it does} hold. This analogous
expression simply involves replacing the original standard errors and
degrees of freedom with their moderated values, \eqref{eqn:dftilde} and \eqref{eqn:stilde}. 
Combining \eqref{eqn:bhatj-t2} with \eqref{eqn:bj-t} then yields
an EBTM problem that can be solved using methods from \cite{Stephens2016}.

\subsection{A two-step strategy}

Putting this all together, we suggest the following two-step strategy for fitting the EBNM model, accounting for uncertainty in estimated standard errors: 
\begin{enumerate}
\item Apply EB shrinkage methods to estimated standard errors $\hat{s}_1,\dots,\hat{s}_p$, using the likelihood \eqref{eqn:a3} and prior \eqref{eqn:seprior}, as in \cite{smyth2004limma}. This yields estimates for $s_0,\df_0$, and subsequently moderated estimates $\tilde{s}_j$ \eqref{eqn:stilde} and degrees of freedom $\tilde{\df}_j$ \eqref{eqn:dftilde}.
\item Apply methods for the EBTM problem \cite[e.g][]{Stephens2016} 
to the estimates $\bhat_j$,
estimated standard errors $\tilde{s}_j$ and degrees of freedom $\tilde{\df}_j$. This yields estimates for $g_\b$ and the posterior distributions $p(\b_j | \bhat_j, \shat_j, \hat{g}_\b)$. 
\end{enumerate}

\subsubsection*{Notes}

\begin{enumerate}
\item Like many two-step procedures, this two-step procedure in not fully efficient: in principle it would be more efficient to {\it jointly} estimate $g_\b, s_0,\df_0$ from $(\bhat_1,\shat_1),\dots,(\bhat_p,\shat_p)$,  rather than first estimate $\s_0,\df_0$ from $\shat_1,\dots,\shat_p$ and then estimate $g_\b$ while fixing the estimates of $\s_0,\df_0$. 
%For example, if $g$ were known to be a point mass on zero, then $\beta_j=0$ for all $j$, and \eqref{eqn:a2} would effectively provide an extra degree of freedom to improve estimation of each $s_j$. 
However in practice, because $p$ is typically large, $s_0,\df_0$ can already be accurately estimated from $\shat_1,\dots,\shat_p$, and in our view the convenience of the two-step procedure greatly outweighs any minor loss of efficiency.
\item The distributional assumption \eqref{eqn:seprior}, which leads to \eqref{eqn:spost}, may seem somewhat restrictive. However, the moderated $t$ statistics from \cite{smyth2004limma} --
which rely on the same assumption -- have been found to be well behaved in practice and are widely used. See \cite{Lu2016,phipson2016} for discussion and assessment of more flexible assumptions.
\item Although assumptions \eqref{eqn:seprior} and \eqref{eqn:a3} are the simplest way to obtain posterior distributions of the form \eqref{eqn:spost}, the form \eqref{eqn:spost} holds more generally. For example, the \emph{voom}  framework \citep{law2014voom} adapts
methods in \cite{smyth2004limma} to deal with the count nature of RNA sequencing data, and involves both accounting for mean-variance relationships and using weighted least squares rather than ordinary least squares. However, it ultimately yields conditional distributions of the form \eqref{eqn:spost}, 
which -- by Lemma 1 -- lead to an EBTM problem for $\b_1,\dots,\b_p$. 
\end{enumerate}

%\subsection{Extensions}

\subsection{Dependence of $\b_j$ on $\shat_j$}

Equation \eqref{eqn:bj-t} assumes that the $\beta_j$ are independent of $\shat_j$. Methods in \cite{Stephens2016} for the EBTM problem can deal with the more general assumption:
\begin{equation}
\b_j /\tilde{\s}_j^\alpha | \shat_j \sim g_\b \in \mathcal{G},
\end{equation}
for any choice of $\alpha \in \mathcal{R}$. 
The choice $\alpha=0$ gives \eqref{eqn:bj-t}.
The choice $\alpha=1$ corresponds to assuming that the moderated $t$ statistics from \cite{smyth2004limma} are independent of $\shat_j$, which in turn leads to the property that EB measures of significance (e.g.~local FDR) are monotonic as the moderated $t$ statistics move away from 0 (and monotonic in the corresponding $p$ values if $g_\b$ is symmetric about 0).
Thus $\alpha=1$ can be thought of as corresponding to the implicit
assumption made when ranking significance by $p$ values from the moderated
$t$ statistics \citep{wakefield2009}. 

Although values of $\alpha$ other than 0 and 1 do not have a straightforward motivation or interpretation, it is straightforward to fit these models, and to estimate $\alpha$ by comparing likelihoods if desired.

\subsection{An {\it ad hoc} strategy that avoids the EBTM problem} \label{sec:adhoc}

The framework outlined above has the advantage of 
being based on clear statistical principles. However, it has the disadvantage that the EBTM problem is often more complex to solve than the EBNM problem. In our numerical studies below we therefore also consider an alternative {\it ad hoc} strategy, which avoids solving the EBTM problem. 

This {\it ad hoc} strategy starts by using the same ideas as above to obtain the estimates $\bhat_j$,
moderated standard errors $\tilde{s}_j$ and degrees of freedom $\tilde{\df}_j$. However, rather than
applying the EBTM methods to these data, we 
convert the problem into a ``normal means" problem, by changing the standard errors. Specifically for each $j$ we
define the ``adjusted standard error'' $s'_j$ to be the value for which the z-score $\bhat_j/s'_j$ results in the same $p$-value (when compared with a standard normal distribution) as from the moderated $t$ test
(comparing $\bhat_j/\tilde{s}_j$ with a $t$ distribution on $\tilde{\df}_j$ degrees of freedom).\footnote{The following R function computes the adjusted standard error from an effect estimate {\tt bhat} and corresponding $p$ value {\tt p}: {\tt pval2se = function(bhat,p)\{z = qnorm(1-p/2); s = abs(bhat/z); return(s)\} }.}
We then use $\bhat_j,s'_j$ as the inputs to an EBNM problem to obtain posterior distributions and shrinkage estimates for
$\beta_j$.

\section{Numerical Studies}

We illustrate our two-stage strategy, and compare it
with the naive strategy, the {\it ad hoc} strategy, and other related methods, using simulations. 
To make our simulated standard errors and test statistics realistic, we base our simulations on real data from an RNA sequencing experiment (RNA-seq data). However, unlike real RNA-seq data, our simulations create data that are independent across genes. In practice RNA-seq data are often
strongly correlated among genes,  and these correlations
can cause severe complications for many analyses methods \citep{leek2007sva}, 
including the Empirical Bayes methods used here \citep{Efron2010,gerard.mouthwash}. 
By removing these correlations here we are comparing methods under idealized 
conditions, and seek to show that even under idealized conditions
the naive approach --
which does not use EB shrinkage of the standard errors -- performs poorly. 
For empirical comparisons of methods on correlated RNA-seq data see
\cite{gerard.mouthwash,lu.thesis}.

\begin{figure}[!ht]
\includegraphics[width=\textwidth]{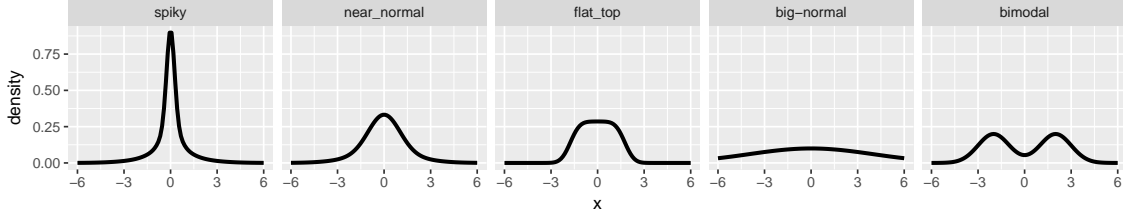}
\caption{Densities of non-zero effects, $g_1$, used in simulations.}
\label{fig:scenarios_dens}
\end{figure}

We perform simulations for two groups, each containing $n$ samples, with $n=2,4,10$ and
$p=10,000$ genes.
The effects are simulated from $\pi_0\delta_0 + (1-\pi_0) g_1$ for various choices of distribution $g_1$ (Figure \ref{fig:scenarios_dens}; Table \ref{table:scenarios}), and then divided by a scaling factor $S_n$ chosen so that power is similar for different $n$ ($S_2=0.125, S_4=0.5, S_{10}=1.5$).
For each combination of $(n,g_1)$, we simulate 50 datasets with $\pi_0$ drawn uniformly from [0,1].

\begin{table}[!ht]
\centering\begin{tabular}{c c } \toprule
Scenario & Alternative distribution, $g_1$  \\ \midrule
spiky & $0.4 N(0,0.25^2) + 0.2 N(0,0.5^2) + 0.2 N(0,1^2), 0.2 N(0,2^2)$\\
near-normal & $2/3 N(0,1^2) + 1/3 N(0,2^2)$ \\
flat-top& $(1/7) [N(-1.5,.5^2) + N(-1,.5^2) + N(-.5,.5^2) +$ \\
 &  $N(0,.5^2) +N(0.5,.5^2)  +N(1.0,.5^2) + N(1.5,.5^2)]$  \\
big-normal & $N(0,4^2)$ \\
bimodal & $0.5 N(-2,1^2) + 0.5 N(2,1^2)$ \\ \bottomrule
\end{tabular}
\caption{Summary of simulation scenarios considered} \label{table:scenarios}
\end{table}

We analyzed each simulated dataset with several methods 
based on the voom-limma (VL) pipeline \citep{law2014voom},
which uses the \texttt{voom} and \texttt{lmFit} functions from the \texttt{limma} R package \citep{ritchie2015}
to obtain estimates $\bhat$ and standard errors $\shat$, along with degrees of freedom $\nu$.
Many of the pipelines also use the \texttt{eBayes} function to obtain moderated standard errors $\tilde{s}$ and moderated degrees of freedom $\tilde{\nu}$, which yield moderated $t$ statistics and corresponding $p$ values.
\begin{itemize}
    \item \va{}: this is the ``naive" approach, which directly feeds the $\bhat,\shat,\nu$ (without variance moderation) into the EBTM solver in the \ash{} function in the {\tt ashr} software \citep{Stephens2016}. As noted above this approach is flawed in principle, and our results show it can also perform poorly in practice.
    \item \vla{} and \vlaa{}: these are our proposed pipelines, which feed the $\bhat$ and moderated standard errors $\tilde{s}$ (and $\tilde{\nu}$) to the \ash{} EBTM solver (with $\alpha=0$ for \vla{} and $\alpha=1$ for \vlaa{}). 
    \item \vlpa{}: this is our ``ad hoc" approach (Section \ref{sec:adhoc}), which converts the EBTM problem into an EBNM problem by computing ``adjusted standard errors'' $s'$, and then applies \ash{} to solve the EBNM problem for $(\bhat,s')$.
    \item \vlq: this is a standard pipeline for controlling FDR in differential expression studies (not based on EB methods): it feeds the $p$ values from the moderated $t$ statistics to the {\tt qvalue} software  \citep{storey2002fdr}, which outputs an estimate for $\pi_0$ and a $q$-value for each test which can be used to control FDR.
\end{itemize}

\paragraph{Estimation of null proportion} \label{sec:fdr_indep}

All of the above methods provide an estimate of the null proportion, $\pi_0$. Obtaining accurate estimates of $\pi_0$ is important for obtaining accurate estimates of FDR: underestimating $\pi_0$ will lead to anti-conservative estimates of FDR, whereas overestimating
 $\pi_0$ will lead to conservative (over-)estimates of FDR, effectively reducing statistical power. 

% An important issue in differential expression analysis is the calibration of FDR. Typically the overall false discovery rate should be controlled under a certain threshold (e.g. 0.05) when declaring significance. For the p-value based methods, it is customary to use multiple testing adjustment procedures (e.g. Benjamini-Hochberg procedure \citep{benjamini1995}, \qvalue{} \citep{storey2002fdr}) to estimate FDR. Here we use the R package \texttt{qvalue} to compute q-values from the p-values of \vl{}. Our method \vla{} directly uses the q-values provided by \texttt{ashr} package, which is computed from the local false discovery rates. Even though both \qvalue{} and \ash{}'s q-values estimate the actual FDR, they have different underlying assumptions: \qvalue{} assumes that alternative genes cannot have effect sizes close to 0 (``Zero Assumption'', denote as ZA), while \ash{} assumes the effect sizes of alternative genes come from an unimodal distribution with mode at 0. As a result, \qvalue{} q-values are expected to be more conservative \ash{} q-values, since \qvalue{} is more likely to treat genes with small effect sizes as null genes.

\begin{figure}[!hbp]
\includegraphics[width=\textwidth]{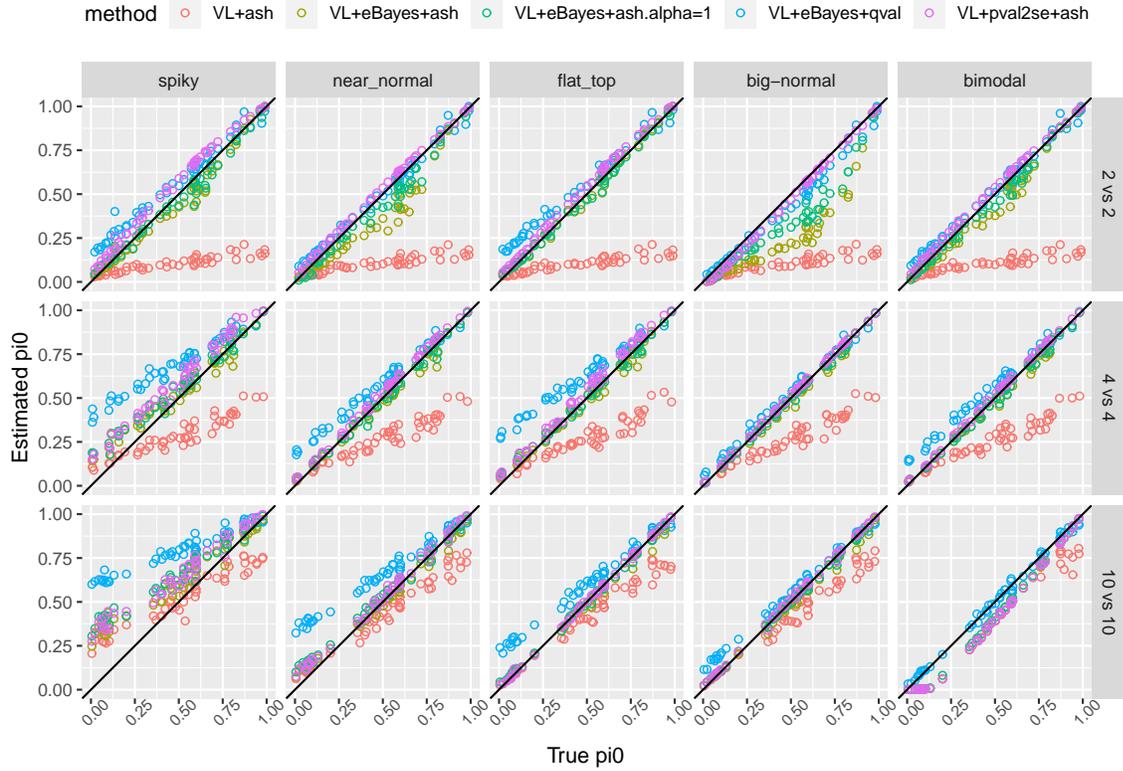}
\caption{Comparison of true and estimated values of $\pi_0$ on simulated data. Generally \va{} is  anti-conservative, often substantially under-estimating $\pi_0$. When the UA holds the other three methods yield conservative (over-)estimates for $\pi_0$, with \vla{}, \vlaa{} and \vlpa{} being less conservative, and hence more accurate. When the UA does not hold (``bimodal" scenario) the \vla{} estimates are slightly anti-conservative.} \label{fig:pi0_indep}
\end{figure}

Figure \ref{fig:pi0_indep} compares the estimated $\pi_0$ with the true $\pi_0$ in our simulations.
The first key observation is that
the naive approach \voom{}+\ash{} can dramatically underestimate $\pi_0$, and cannot be recommended.
All other approaches generally provide reasonable (conservative) estimates of $\pi_0$, with
the \ash{}-based approaches producing more accurate (less conservative) estimates than those from \qvalue{}. This improved accuracy comes from the additional assumption made by the EB approach in \ash{}, that the effects are unimodal \citep{ Stephens2016}. The results are reasonably robust to this assumption, but estimates of $\pi_0$ can be anti-conservative in the bimodal scenario (just as in \cite{Stephens2016}).

% Note that data are simulated under the $\alpha=0$ model (effect sizes are exchangeable), but \qvalue{} assumes the $\alpha=1$ model (z-scores are exchangeable). 

\paragraph{Assessment of FDR control and power}

\begin{figure}[!hbp]
\includegraphics[width=\textwidth]{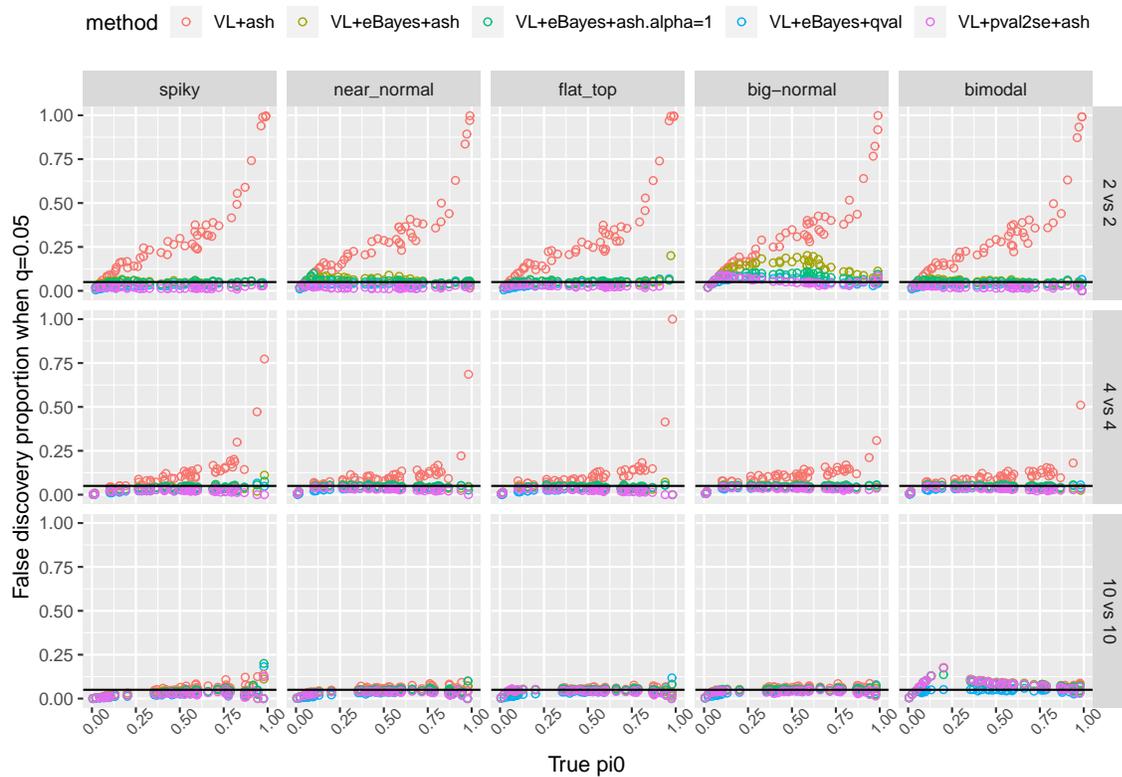}
\caption{Comparison of empirical false discovery proportions (FDP), at $q$-value $<0.05$, on simulated data . Generally the naive method, \va{}, is  anti-conservative, failing to control FDP $<0.05$. In contrast, other methods generally control FDP near or under 0.05,  although \vla{} is slightly anti-conservative in the ``big-normal" scenario with small sample size ($2$ vs $2$). } \label{fig:fdp_indep}
\end{figure}

\begin{figure}[!hbp]
\includegraphics[width=\textwidth]{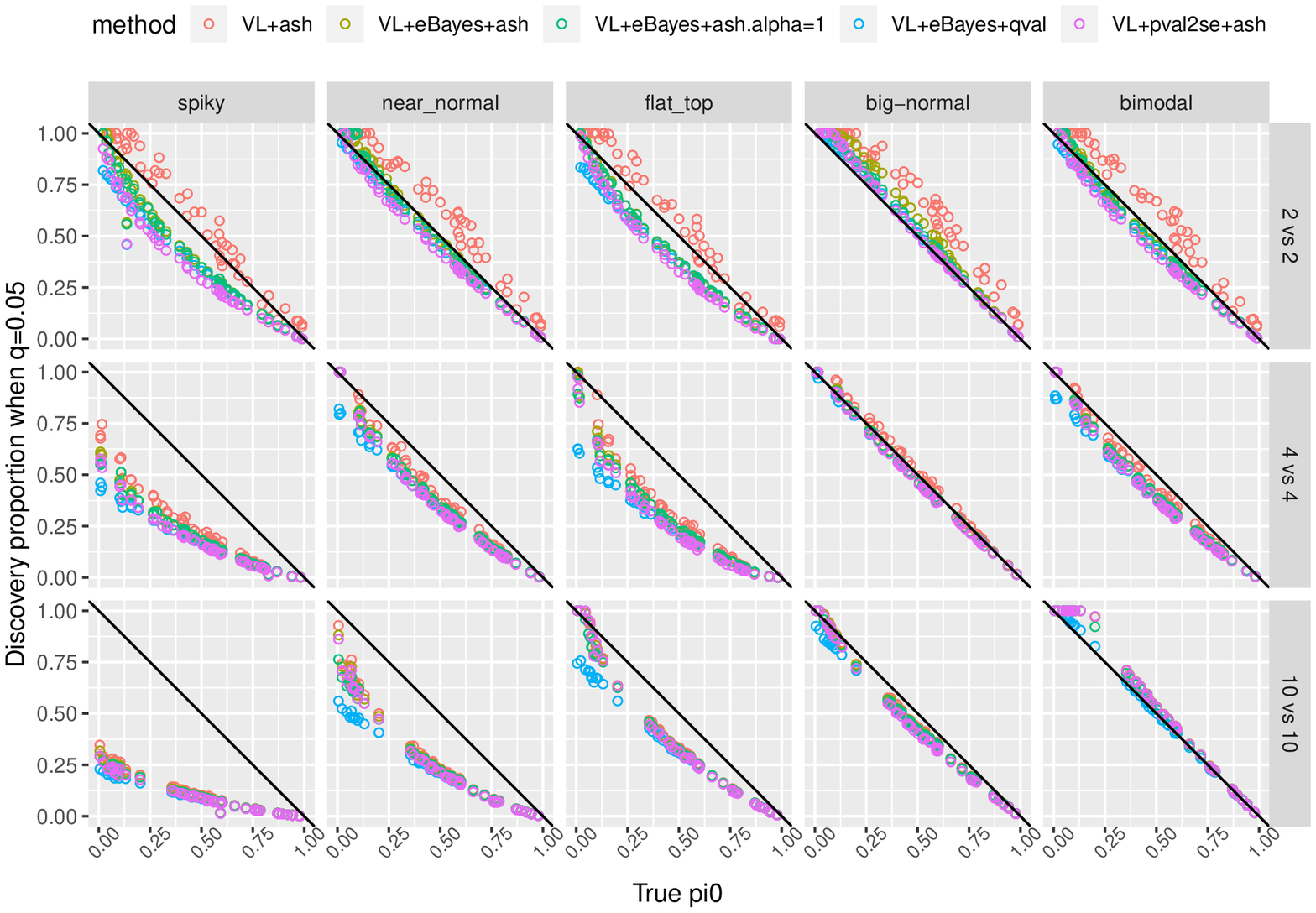}
\caption{Comparison of proportion of discoveries (``power"), at $q$-value $<0.05$, on simulated data. Typically \vla{} and \vlaa{} have more discoveries than \vlq{}, while controlling FDP (Figure \ref{fig:fdp_indep}).} \label{fig:dp_indep}
\end{figure}

Figures \ref{fig:fdp_indep}  assesses how well each method controls FDR in our simulations (at nominal level 0.05), and Figure  \ref{fig:dp_indep} shows the corresponding power (proportion of true effects declared significant). The naive method completely fails to control FDR for very small sample sizes (2 vs 2 or 4 vs 4). Other methods perform generally well at controlling FDR, although there is some lack of FDR control of \ash{}-based methods in the bimodal scenario. The \ash{}-based methods are slightly more powerful than \qvalue{} because of the less conservative estimates of $\pi_0$.

%\begin{figure}[!hbp]
%\includegraphics[width=\textwidth]{./plots/fdp_qval_indep.eps}
%\caption{Comparison of actual false discovery proportions on simulated data  if declaring genes with $q$-values under 0.05 as positives. \vlq{} is generally able to control the false discovery proportion under 0.05 regardless of sample size, while \deseq{} and \edger{} are typically anti-conservative.} \label{fig:fdp_indep}
%\end{figure}

%\begin{figure}[!hbp]
%\includegraphics[width=\textwidth]{./plots/fsp_indep.eps}
%\caption{Comparison of actual false sign proportions on simulated data  if declaring genes with $s$-values under 0.05 as positives. \vlq{} and \vla{} are generally able to control the false sign proportion under 0.05 regardless of sample size. \vla{} can be slightly anti-conservative when the UA does not hold (``bimodal" scenario) and $\pi_0$ is less than 0.5. } \label{fig:fsp_indep}
%\end{figure}

\paragraph{Effect estimates} \label{sec:ee_indep}

One attractive feature of the EB approach to multiple testing is that it provides not
only estimates of FDR, but also shrinkage estimates of effect sizes. To compare the accuracy  of the shrinkage estimates with the original (unshrunk \vl{}) estimates
we compute the relative root mean squared error (RRMSE) for each method as the ratio of the method's RMSE and the baseline RMSE for the original estimates. Here  $\text{RMSE}:= \sqrt{\sum_j (\tilde{\beta}_j-\beta_j)^2}$. 
 
\begin{figure}[!hbp]
\includegraphics[width=\textwidth]{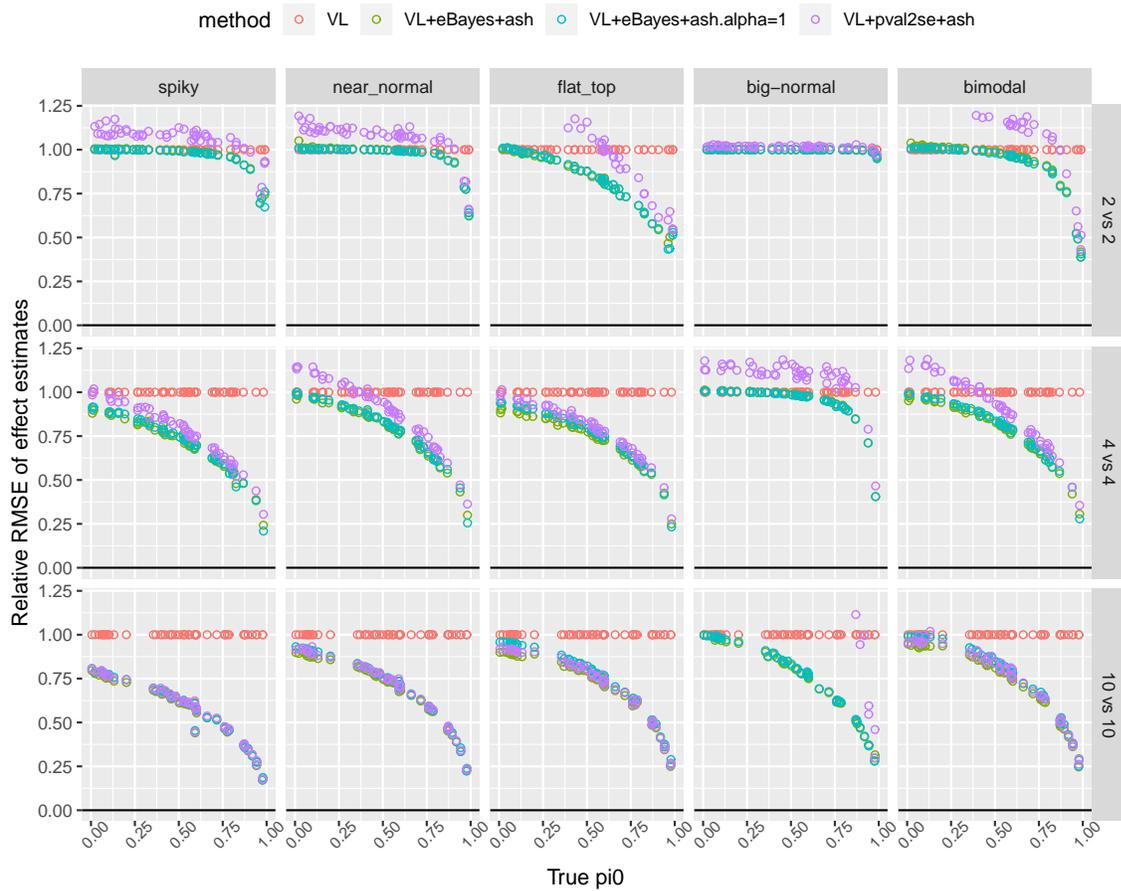}
\caption{Comparison of RRMSE (relative root mean squared error) of effect estimates on simulated data . To compute RRMSE we compute RMSE of \emph{VL} as the baseline level, and divide the RMSE of each method by this baseline. Thus by definition RRMSE of \emph{VL} is 1. \vla{} is more accurate (RRMSE<1) in all scenarios, especially when $\pi_0$ is close 1. The ad hoc approach \vlpa{} performs less well for small sample sizes (although similar to \vla{} for $10$ vs $10$, except for ``big-normal" scenario).} \label{fig:rmse_indep}
\end{figure}

The results (Figure \ref{fig:rmse_indep}) demonstrate the expected benefits of shrinkage estimation:
the shrunken estimates from \vla{} (whether $\alpha=0$ or $1$) are consistently better than the original unshrunk estimates. The improvement on the baseline RMSE is up to 90\% in settings where most effects are null, where the benefits of shrinkage are strongest.

 %\deseq{} can decrease the baseline RMSE by up to 40\% in ``spiky'', ''near normal'' and ''flat top'' scenarios,
 
\paragraph{Calibration of posterior intervals}

In addition to shrinkage point estimates of $\beta$, the EB approach also provides ``shrunk" interval estimates. \cite{Stephens2016} used simulations to show
that, under idealized conditions (with known standard errors), these interval estimates not only have good coverage properties on average, but also ``post-selection": that is, even if we focus only on significant effects, the coverage of the EB credible intervals is good. This property is difficult to obtain in other ways.

Here we repeat this coverage assessment in the case of estimated standard errors.
Table \ref{tab:coverage_indep} shows the coverage rates of 95\% lower credible bounds for the effects, split into all observations (a), significant negative discoveries (b) and significant positive discoveries (c). Note that for significant negative discoveries (b), the lower credible bound is bounding how ``large" the effect is (in absolute value), whereas
for positive discoveries (c) it is bounding
how close to 0 it can be. In general coverage rates are satisfactory, with the most prominent exception being the case $N=2$ in (b), where coverage rates are often much lower than the nominal 95\%.  
This says that the method is ``over-shrinking" the significant effects towards zero in this case, probably due to underestimating the length of the tail of the effects. In low-signal situations some level of over-shrinkage may be inevitable if we want to maintain conservative behaviour (i.e avoid under-shrinkage); thus it is unclear to what extent
this behavior could be improved on.

%\begin{figure}[!hbp]
%\includegraphics[width=\textwidth]{./plots/fsp_indep.eps}
%\caption{Comparison of actual false sign proportions on simulated data  if declaring genes with $s$-values under 0.05 as positives. \vla{} always controls the actual false sign proportion under 0.05.} \label{fig:fsp_indep}
%\end{figure}

\begin{table}[h!]
\begin{subtable}{\textwidth}
	\centering\input{./tables/coverage_all.tex}

	\caption{All observations. Coverage rates are generally satisfactory, except for the big-normal scenario case when $N$=2.}
\end{subtable}

\begin{subtable}{\textwidth}
\centering\input{./tables/coverage_neg.tex}

\caption{``Significant" negative discoveries. Coverage rates are generally satisfactory when $N=10$ and for $N=4$ (except for big-normal scenario), but are generally poor for $N=2$,
suggesting over-shrinkage (underestimating the length of the tail of $g$) in this setting.}
\end{subtable}

\begin{subtable}{\textwidth}
\centering\input{./tables/coverage_pos.tex}

\caption{``Significant" positive discoveries. Coverage rates are generally satisfactory. }
\end{subtable}

\caption{Table of empirical coverage for nominal 95\% lower credible bounds for method \vla{} on simulated data .} \label{tab:coverage_indep}
\end{table}

\subsection{Discussion}

In summary, we have shown how EB analysis of normal means with estimated standard errors can be satisfactorily solved by performing an EB analysis 
of the ``$t$-means" problem (EBTM), but only
after applying EB methods to the estimated standard errors themselves to 
obtain moderated estimates of the standard errors (and associated degrees of freedom). 

Our numerical results also show that a simpler {\it ad hoc} approach, \vlpa{} in the Figures,
which avoids solving the more complex EBTM problem by instead adjusting the standard errors and solving an EBNM problem, can work adequately to control false discovery rates. However, it performs less well in estimation accuracy than the more principled approaches based on solving the EBTM problem. 

Code used to obtain the numerical results presented
here is available at
\url{http://doi.org/10.5281/zenodo.2547022}.

\clearpage 

\bibliography{document}
\bibliographystyle{chicago}
%\nocite{*}

\clearpage

\appendix

\section{Simulation details} \label{sec:simscheme}

The following simulation scheme is designed to
create realistic count datasets that mimic the
structure of real RNA-seq data, making distributional assumptions only sparingly.
\begin{enumerate}
\item Start with real RNA-seq data on $J$ genes in $N$ samples.
\item  Create a ``null" data set containing two groups ($A$ and $B$), of sizes $n_A,n_B$, by randomly sampling (without replacement) $n_A$ samples for group $A$ and $n_B$ samples for group $B$. 
Because the assignment of samples to the two groups is random, this is a null dataset by construction. Let $C_{ji}$ denote the read count for gene $j$ and sample $i$.
\item Randomly select $J(1-\pi_0)$ genes as ``alternative genes'', and generate their effects ($\log_2$-fold-change between groups) $\beta_j$'s from a specified ``effect distribution" $g_1$.
\item For these alternative genes, if $\beta_j>0$ (so group $B$ should be more highly expressed), we use Poisson thinning to achieve the desired fold-change $2^{\beta_j}$ i.e. thin the read counts in group $A$ as follows:
\begin{equation}
C_{ji}^* \sim Binomial(C_{ji}, 2^{-\beta_j}), \quad \forall i\in A.
\end{equation}
Similarly if $\beta_j<0$, thin the read counts in group $B$:
\begin{equation}
C_{ji}^* \sim Binomial(C_{ji}, 2^{-\beta_j}), \quad \forall i\in B.
\end{equation}
Replacing $C_{ji}$ by $C_{ji}^*$ will result in a new RNA-seq dataset, where the true effects follow $\pi_0 \delta_0 + (1-\pi_0) g_1$.
\end{enumerate}

\subsubsection*{Simulations }  \label{sec:sim_indep}

The above simulation scheme, which we developed during
our work on this project, was used by \cite{gerard.mouthwash} to generate realistic simulated RNA-seq datasets with a desired effect distributions, while still preserving most of the structure (correlation, magnitude, etc) of the actual RNA-seq data. Unfortunately correlations among genes create substantial complications for many analysis methods \citep{leek2007sva}, including ours; see \cite{gerard.mouthwash} for extensive discussion and further references. To avoid these
complications here we modify this scheme to {\it remove
correlations between genes}. Specifically we modify step 2 
to randomly select the $n_A$ and $n_B$ samples for groups
$A$ and $B$ {\it independently at each gene}.
This modification ensures that the simulated null data at each gene are independent.

While this modification makes the simulations
unrepresentative of typical RNA-seq experiments
(since real data are typically correlated across genes), it allows us to study the behaviour of methods under idealized situations, which is helpful for understanding
the main conceptual contribution of our work here.
Results of our methods on the more realistic simulations
with correlations intact are given in \cite{lu.thesis}.

Our simulations here used RNA-seq data from liver tissue
samples distributed by the Genotype-Tissue Expression (GTEx) project \citep{gtex}. These data (GTEx V6 dbGaP accession phs000424.v6.p1, release date: Oct 19, 2015, \href{http://www.gtexportal.org/home/}{http://www.gtexportal.org/home/}) contained data on 119 samples, and we restricted simulations to the 10,000 top expressed genes. 

\end{document}

%% file: tables/coverage_all.tex
% latex table generated in R 3.5.1 by xtable 1.8-3 package
% Thu Jan 17 11:44:00 2019
\begin{tabular}{rrrrrr}
  \toprule  & big-normal & bimodal & flat-top & near-normal & spiky \\ 
  \midrule N=2 & 0.80 & 0.93 & 0.96 & 0.91 & 0.93 \\ 
  N=4 & 0.93 & 0.96 & 0.96 & 0.96 & 0.95 \\ 
  N=10 & 0.97 & 0.97 & 0.96 & 0.96 & 0.95 \\ 
   \bottomrule \end{tabular}

%% file: tables/coverage_neg.tex
% latex table generated in R 3.5.1 by xtable 1.8-3 package
% Thu Jan 17 11:43:59 2019
\begin{tabular}{rrrrrr}
  \toprule  & big-normal & bimodal & flat-top & near-normal & spiky \\ 
  \midrule N=2 & 0.23 & 0.74 & 0.94 & 0.65 & 0.74 \\ 
  N=4 & 0.76 & 0.95 & 0.94 & 0.95 & 0.94 \\ 
  N=10 & 0.95 & 0.94 & 0.94 & 0.94 & 0.93 \\ 
   \bottomrule \end{tabular}

%% file: tables/coverage_pos.tex
% latex table generated in R 3.5.1 by xtable 1.8-3 package
% Thu Jan 17 11:43:59 2019
\begin{tabular}{rrrrrr}
  \toprule  & big-normal & bimodal & flat-top & near-normal & spiky \\ 
  \midrule N=2 & 0.98 & 0.96 & 0.96 & 0.96 & 0.96 \\ 
  N=4 & 0.96 & 0.96 & 0.96 & 0.96 & 0.96 \\ 
  N=10 & 0.95 & 0.96 & 0.96 & 0.96 & 0.96 \\ 
   \bottomrule \end{tabular}